\documentclass[preprint,showkeys,aps,prb]{revtex4}
\usepackage[dvips]{graphicx}

\begin{document}

\title{The 2017 SNOOK PRIZES in\\
 Computational Statistical Mechanics}

\author{
Wm. G. Hoover and Carol G. Hoover               \\
Ruby Valley Research Institute                  \\
Highway Contract 60, Box 601                    \\
Ruby Valley, Nevada 89833                       \\
}

\date{\today}

\keywords{Lyapunov Instability, Reversibility, Second Law of Thermodynamics, Time-Symmetry Breaking}

\vspace{0.1cm}

\begin{abstract}

The 2017 Snook Prize has been awarded to Kenichiro Aoki for his exploration of
chaos in Hamiltonian $\phi^4$ models.  His work addresses
symmetries, thermalization, and Lyapunov instabilities in few-particle dynamical
systems.  A companion paper by Timo Hofmann and Jochen Merker is devoted to the
exploration of generalized H\'enon-Heiles models and has been selected for Honorable
Mention in the Snook-Prize competition.

\end{abstract}

\maketitle

\section{Introduction}

The annual Snook Prize awards honor our late Australian colleague Ian Snook ( 1945-2013 )
and his contributions to computational statistical mechanics during his 40-year
tenure at the Royal Melbourne Institute of Technology.  The Snook Prize Problem for 2017
was to explore and discuss the uniqueness of the chaotic sea, the spatial variation of
local small-system values of the kinetic temperature at equilibrium,
the time required for Lyapunov-exponent pairing and its dependence on the integrator,
as well as the possibility of the long-time {\it absence} of exponent pairing\cite{b1}.
The Lyapunov exponents, two for each Hamiltonian degree of freedom, characterize the
strength of chaos in classical dynamical systems.

Kenichiro Aoki took the Snook Prize challenge seriously and explored the behavior of
$\phi^4$-model exponent pairs, including the dependence of pairing on energy,
symmetry, and system size. In his work\cite{b2} Aoki asked whether or not chaotic
ergodicity implies a uniform temperature distribution in phase space, and whether or
not disparities in kinetic temperature could persist for long in isoenergetic chaotic
Hamiltonian steady states,
$$
mkT_i = \langle \ p_i^2 \ \rangle \stackrel{?}{=}
\langle \ p_j^2 \ \rangle = mkT_j \ .
$$
Persistent temperature differences might well  appear to enable violations of the Second Law
of Thermodynamics through the spontaneous generation of enduring temperature gradients.
Aoki demonstrated the existence
of microcanonical temperature differences along with a lack of long-time exponent
pairing. He found faster pairing for exponents at higher temperatures.  He explored
the effects of symmetries on the dynamics of anharmonic chains and on their exponents.
He pointed out the short-term dependence of the exponents on the ordering of the
variables using the Gram-Schmidt orthonormalization algorithm.  We have judged his work
worthy of the 2017 Snook Prize, awarded in April 2018.

\section{The $\phi^4$ Model Generates Chaotic Dynamical Systems}
\subsection{Chaos with Just Two One-Dimensional Bodies}
Kenichiro Aoki and Dimitri Kusnezov\cite{b3} pioneered investigations of the $\phi^4$
model as a prototypical ideal material supporting Fourier heat conduction.  The model
augments the harmonic chain with the addition of attractive quartic tethering potentials which
bind the particles to individual lattice sites. The
simplest interesting chaotic case invoves two masses and two springs.  It is shown at the top of
{\bf Figure 1} :
$$
{\cal H} = (1/4)(q_1^4+q_2^4) + (1/2)[ \ p_1^2+p_2^2 + q_1^2 +(q_1-q_2)^2 \ ] \ .
$$
For convenience the masses and spring constants have all been set equal to unity.
The rest positions of both particles define the coordinate origins,
$q_1 = 0 \ ; \ q_2 = 0$. Particle 1 is linked to a fixed boundary to its left with
$\phi_1 = (q_1^2/2)$. Particle 1 is additionally linked to Particle 2 at its right with
$\phi_{12} = (q_1-q_2)^2/2$. {\bf Figure 1} shows the momentum correlation for the two
Particles for four different initial conditions, all of them with total energy ${\cal H} = 6$.

\noindent
\begin{figure}[h]
\includegraphics[width=3.0in,angle=0,bb= 16 71 600 726]{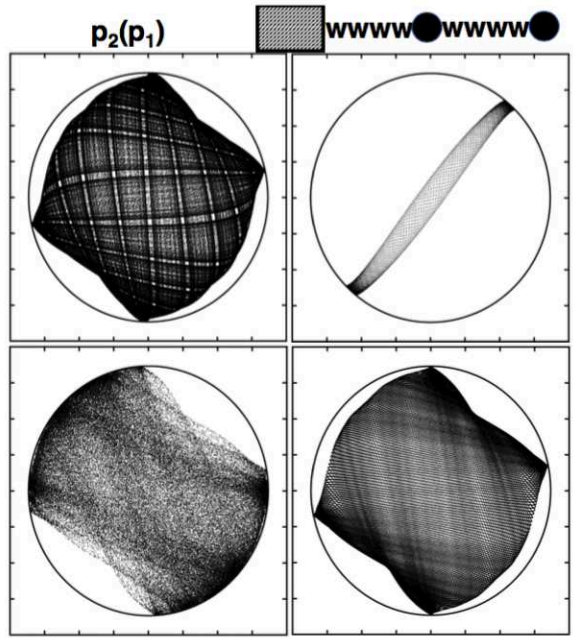}
\caption{The top row shows $p_2(p_1)$ projections for initial values $\{ \ p^2 \ \} =
(11.6,0.4)$ at the left and $(6,6)$ at right.  The bottom row shows $p_2(p_1)$ projections
for initial squared momenta of (12,0) at left and (0,12) at right.  Runge-Kutta fourth-order
integration with $dt = 0.001$ and ${\cal H} = 6$ are used in Figures 1-3.  The bottom left
projection represents chaos, with a Lyapunov exponent of 0.05.  The other three do not. It
appears that the combination (11.4,0.6) {\it is} chaotic with a Lyapunov exponent of order
0.003.  The circles in {\bf Figure 1} represent the maximum momenta, $p^2 = 12$.
}
\end{figure}

\noindent
\begin{figure}[h]
\includegraphics[width=3.0in,angle=-90,bb=35 37 580 758]{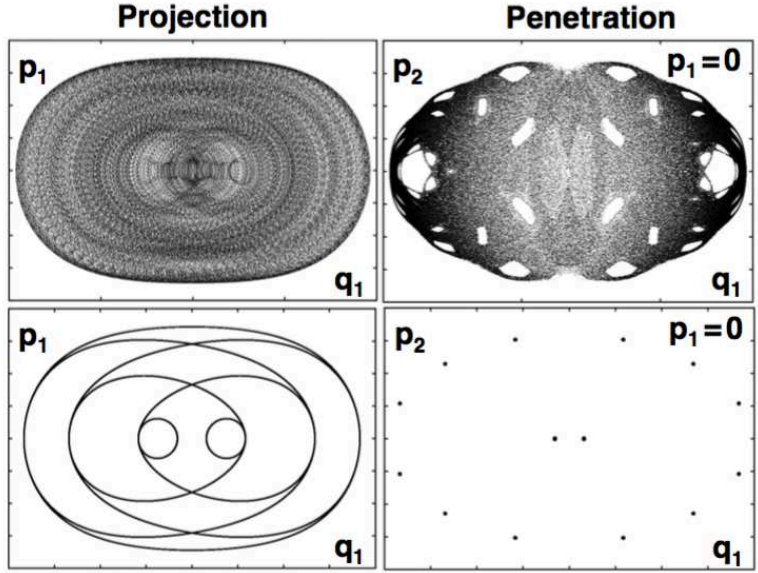}
\caption{
The left column shows phase-plane projections $p_1(q_1)$ and the right column $p_2(q_1)$ shows
Poincar\'e sections $p_2(q_1)$ with $p_1=0$. The initial momenta for the top row are $\{ \ p^2 \ \}
= (12,0)$ and for the bottom row $(11.85,0.15)$. The top row is chaotic with $\lambda_1 = 0.05$
while the bottom row is close to a periodic orbit with $\lambda_1 = 0$. The top-row section suggests a
``fat fractal'' with infinitely-many tori threading through its perforations.
}
\end{figure}

\noindent
\begin{figure}[h]
\includegraphics[width=3.0in,angle=-90,bb=51 65 556 730]{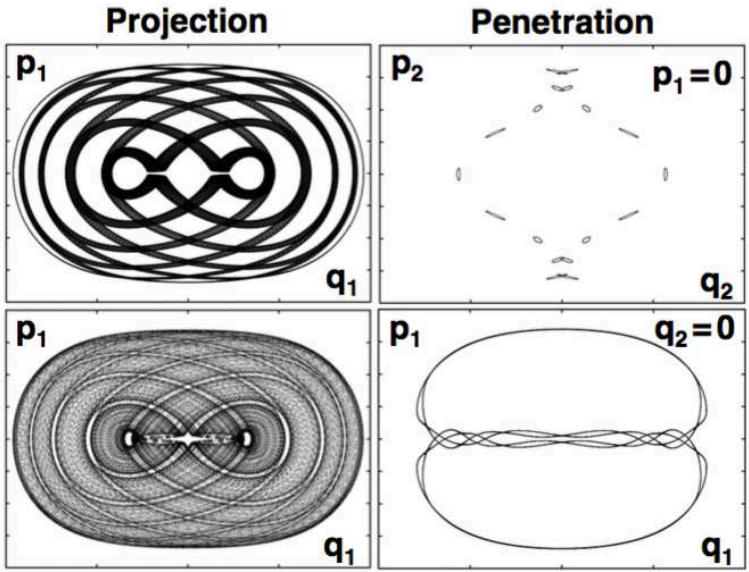}
\caption{
Here the top row initial momenta are $\{ \ p^2 \ \} = (11.5,0.5)$ and the motion is regular,
with $\lambda_1 = 0$.  The bottom row momenta are $(11.4,0.6)$ where chaos or its lack are
still problematic, a region meriting more study. Here the left panels are projections while
the right ones show Poincar\'e penetrations.
}
\end{figure}

Aoki goes well beyond this interesting two-body problem.  He considers a variety of $\phi^4$
chain problems in his provocative paper, considering the effects of symmetry and chaos on
equilibrium temperature distributions, Lyapunov instability, and the pairing of the Lyapunov
exponents. At present it is unknown whether or not the spectra become paired for {\it all},
or maybe ``almost all'' initial conditions. The $\phi^4$ model provides a surprisingly rich
testbed for investigating Lyapunov instability.

Poincar\'e sections, two-dimensional cuts through the three-dimensional phase-density in its
four-dimensional embedding space are shown at the right in {\bf Figure 2}.  The upper right
section suggests a distribution
resembling a ball of yarn with holes here and there, with these holes filled with periodic
tori.  Although far from ergodic the one-dimensional trajectory certainly explores most of the
${\cal H}=6$ region for this simplest of initial conditions where a single momentum variable
carries {\it all} of the energy.  Despite its two-spring simplicity the model's quartic tethers
are enough to bring out all the general features associated with Hamitonian chaos.

The Sections of {\bf Figure 3} hint at the ``chains-of-islands'' features typical of the boundaries separating
regular and chaotic regions in sections of Hamiltonian chaos.  A series of squared momenta
in the neighborhood of (11.5,0.5) to 11.4,0.4) would make an interesting research
investigation. Aoki also investigated a sightly more complex model with four degrees of freedom
rather than just two.  Let us consider it next.

\subsection{Chaos with Four Bodies Subject to Periodic Boundary Conditions}

A four-body periodic chain of $\phi^4$ particles has the Hamiltonian
$$
{\cal H} = (1/4)(q_1^4+q_2^4+q_3^4+q_4^4) + (1/2)(p_1^2+p_2^2+p_3^2+p_4^2) + 
$$
$$
(1/2)[ \ (q_1-q_2)^2+(q_2-q_3)^2+(q_3-q_4)^2+(q_4-q_1)^2 \ ] \ .
$$
With the implied periodic boundary conditions the special initial conditions ,
$$
q_1=q_2=q_3=q_4 = 0 \ ; \ p_1 = p_2 = -p_3 = p_4 = 2 \ , 
$$
generate an interesting chaotic solution. It is symmetric. Consequently Particles 2
and 4 obey {\it identical} equations of motion. If as here the initial conditions
have proper symmetry, identical dynamics results, with $q_2(t) = q_4(t)$ for all time. The
motion corresponding to these conditions is chaotic.  It likely generates a
five-dimensional isoenergetic and chaotic ``fat fractal'', in either six-dimensional
or eight-dimensional phase space, $\Gamma = \{ q,p \}$. Here the $\{q,p\}$ are the
Cartesian coordinate-momentum pairs describing the system dynamics.

 {\bf Figure 4} demonstrates the chaotic nature of that symmetric motion.  It displays the
convergence process $\langle \ \lambda_1(t) \ \rangle_{t \rightarrow \infty} \longrightarrow
\lambda_1$ . The maximum Lyapunov exponent $\langle \ \lambda_1(t) \ \rangle$ is calculated
by following {\it two} neighboring phase-space trajectories, the ``reference'' and a nearby ``satellite''
in either six or eight-dimensional phase space.  The satellite trajectory is kept close to the reference
by a rescaling of its phase-space separation from that reference  at
the end of every timestep :
$$
| \ \Gamma_r(t+dt) - \Gamma_s(t+dt) \ | \longrightarrow \delta \ .
$$
This condition is implemented by rescaling the separation to the fixed length
$\delta$ \ :
$$
\Gamma_s(t+dt) \longrightarrow \Gamma_r(t+dt) +
\delta \frac{ \ \Gamma_s(t+dt) - \Gamma_r(t+dt) \ }{| \ \Gamma_s(t+dt) - \Gamma_r(t+dt) \ |} \ .
$$
The ``local'' ``largest'' Lyapunov exponent, $\lambda_1(t)$ for that timestep, is
given in terms of the phase-space offsets before and after the rescaling operation :
$$
| \ \delta \ | \propto e^{\lambda t} \longrightarrow
\lambda_1(t) \equiv (1/dt)\ln[ \ {\delta}_{\rm before}/\delta_{\rm after} \ ] \ .
$$
Note well that what is here called ``largest'' is so only as a time average.  The
``largest'' can at times actually be the {\it smallest} !

{\bf Figure 4} shows both six-dimensional and eight-dimensional cumulative time
averages of the local largest Lyapunov exponent, as computed with fourth- and
fifth-order Runge-Kutta integrators.  The chaos of the $\phi^4$ dynamics guarantees
that different integrators eventually generate different trajectory pairs and
different local Lyapunov exponents. But the ``global'' long-time averages of the
Lyapunov exponents are in statistical agreement with each other for any properly 
convergent integrator.

\noindent
\begin{figure}[h]
\includegraphics[width=2.5in,angle=90,bb=142 229 482 697]{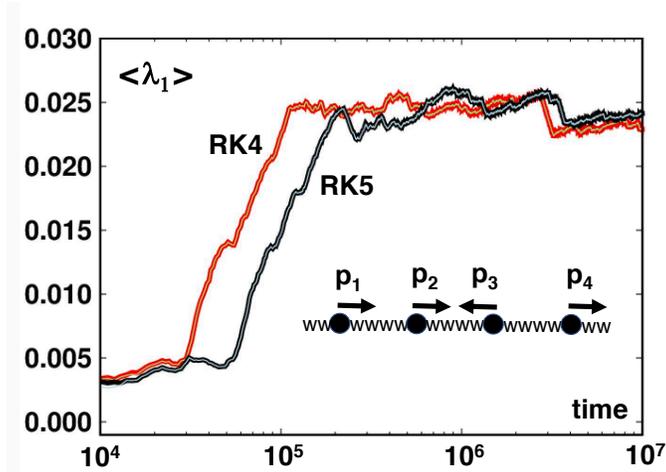}
\caption{
The cumulative ``largest'' Lyapunov exponent describing the four-particle
chaos of a $\phi^4$ system with initial velocities $\{2,2,-2,2\}$ as
computed in six-dimensional (thin lines) and eight-dimensional spaces ( thick lines )
using fourth-order and fifth-order Runge-Kutta integrators. For these simulations,
with 10,000,000,000 timesteps $dt= 0.001$, the reference trajectories agree precisely
while the satellite trajectories show real, but negligibly small, differences.\\
}
\end{figure}

Unlike many simple one-dimensional models augmenting harmonic chains the $\phi^4$ chains
are tethered to fixed lattice sites by a quartic potential :
$$
\{ \ \ddot q_i = q_{i-1} - 2q_i + q_{i+1} - q_i^3 \ \} \
[ \ {\rm Newtonian} \ \phi^4 \ {\rm Model} \ ] \ .
$$
The tethers kill the ballistic transmission of energy.  Over wide ranges of energy and
temperature $\phi^4$ chains exhibit Lyapunov instability ( exponential growth of small
perturbations ) and Fourier conductivity, even for just a few particles.  The initial
conditions are not crucial so long as we choose them from a unique chaotic sea.  With
uniqueness {\it long-time} averages, either at or away from equilibrium conform to
well-defined long-time averages.  Until now investigations of small-system chaos have
typically found a single chaotic sea for a given energy.  This apparent simplicity
could be misleading. In Aoki's two-body $\phi^4$ model, pictured in {\bf Figures 1-3},
$\lambda_1$ is 0.05 with the
initial squared momenta $(12.0,0.0)$. For moderate run lengths at the same energy the initial
momenta (11.4,0.6) lead instead to a smaller Lyapunov constant, $\simeq 0.003$.  In their
contribution Hofmann and Merker likewise suggest that the chaotic sea is not unique in
a generalized H\'enon-Heiles model with two degrees of freedom.

\subsection{Fourier Heat Flow in Chaotic $\phi^4$ Systems}
Hot and cold boundaries can be introduced into chaotic systems by adding
temperature-dependent ``thermostat'' forces directing the kinetic temperatures
of two or more particles' mean values to their individual values
$\{T_i\}$ :
$$
\ddot q_i = F \longrightarrow \ddot q_i = F - \zeta \dot q_i \ {\rm with}
\ \dot \zeta = \dot q_i^2 - T_i \ 
[ \ {\rm Nos\acute{e}-Hoover} \ \phi^4 \ {\rm Model} \ ] \ .   
$$
The heat current that results reduces the dimensionality of the phase-space
distribution from the equilibrium value of $2N$ for $N$ degrees of freedom
by an amount roughly quadratic in the heat current.  Reductions of the order
of 50$\%$ have been obtained in steady-state heat flow simulations with a
few dozen particles, one hot, one cold, and the rest purely Newtonian\cite{b4}.

These results follow from measurements of the {\it spectrum} of Lyapunov exponents\cite{b4}. 
$\lambda_1(t)$ quantifies the local rate at which nearby pairs of phase-space
trajectories separate. $\lambda_1+\lambda_2$ quantifies the rate at which {\it triangles}
of phase-space trajectories grow or shrink, and $\lambda_1 + \lambda_2 + \lambda_3$
does this for tetrahedra. When such a sum of $n$ Lyapunov exponents has a positive
average, overall growth of $n$-dimensional phase volumes occurs, and at the
rate $\sum \lambda_{i\leq n}$.  A negative sum indicates shrinkage onto a strange
attractor with a dimensionality less than that of the full phase space.

In Hamiltonian systems growth and shrinkage necessarily cancel due to the
time-reversibility of the motion equations.  Nonequilibrium  simulations with
thermostats are typically time-reversible too. Both $\{ \ \ddot q \ \}$ and the
thermostat forces $\{ \ -\zeta \dot q \ \}$ are even functions of the time.
Nevertheless, in practice the greater abundance of compressive phase-space states
causes symmetry breaking.  The invariable shrinkage of the comoving phase volume onto
a nonintegral-dimensional ``fractal'' strange attractor results.  In the nonequilibrium
case the spectrum of exponents is {\it always} shifted toward more negative values.
This is the heat-based mechanical explanation of the Second Law of Thermodynamics.

\section{H\'enon-Heiles' Model}
A more mathematically oriented paper by Timo Hofmann and Jochen Merker\cite{b5}
took Honorable Mention in the Snook Prize competition with their study of the
H\'enon-Heiles model in a four-dimensional Hamiltonian phase space :
$$
{\cal H} = (1/2)(p_x^2 + p_y^2 + x^2 + y^2) + x^2y - (y^3/3) \
[ \ {\rm H\acute{e}non-Heiles} \ {\rm Model} \ ] \ .
$$
H\'enon-Heiles models are unsuited to conductivity or to manybody problems
but still have considerable interest.  Hofmann and Merker made a plausible argument
for the coexistence of two chaotic seas for a generalized H\'enon-Heiles
model augmented by three quartic, three quintic, and four sextic terms in
the four-variable generalized Hamiltonian.

Hofmann and Merker compared three versions of local Lyapunov exponents for their
model and concluded that the Gram-Schmidt local exponents are not necessarily
paired because their values depend upon the initial conditions.  We decided to
award their work an Honorable Mention in view of its interest and their
exploration of a problem area quite different to Aoki's.  We congratulate all
three men.

\section{Epilogue and Moral}
For Carol and me it was a challenge to reproduce some of the work of all
the Snook Prize entries.  It often takes real effort to dissipate the
uncertainty characteristic of numerical chaos work. But by using reproducible
``random number generators'', straightforward integrators, and careful
descriptions of {\it all} the necessary initial conditions it is still possible to
describe reproducible results, the {\it sine qua non} of physics.  We are very
grateful to Ken Aoki, Timo Hofmann, and Jochen Merker for their useful and
interesting contributions, as well as the welcome input from several anonymous referees
and colleagues.

\section{Acknowledgments}
We are particularly grateful to Kris Wojciechowski for stimulating and
overseeing this work and to Giancarlo Benettin, Carl Dettmann, Puneet Patra,
Harald Posch, and Roger Samelson for their patient comments and suggestions
regarding aspects of last year's Snook Prizes.  We very much appreciate the
Institute of Bioorganic Chemistry of the Polish Academy of Sciences and the
Pozna\'n Supercomputing and Networking Center for their continued support of
the Ian Snook Additional Prize.

\pagebreak

\end{document}